%% file: elsdoc.tex
\documentclass[1p]{elsarticle}
\usepackage{lineno}
\usepackage{xcolor}
\usepackage{amsmath}
\usepackage{graphicx}
\usepackage{caption}
\usepackage{subcaption}
\usepackage{float}
\usepackage{multicol}

\begin{document}

\title{The Highly-Granular Time-of-Flight Neutron Detector for the BM@N experiment}

\author[inr]{S.\,Morozov\corref{cor1}}
\ead{morozovs@inr.ru}
\author[hse]{V.\,Bocharnikov}
\author[inr]{D.\,Finogeev}
\author[inr]{M.\,Golubeva}
\author[inr]{F.\,Guber}
\author[inr]{A.\,Izvestnyy}
\author[inr]{N.\,Karpushkin}
\author[inr]{A.\,Makhnev}
\author[inr]{P.\,Parfenov}
\author[inr]{D.\,Serebryakov}
\author[inr]{A.\,Shabanov}
\author[inr]{V.\,Volkov}
\author[inr]{A.\,Zubankov}

\cortext[cor1]{Corresponding author}

\address[inr]{Institute for Nuclear Research of the Russian Academy of Sciences,\\60-letiya Oktyabrya prospekt 7a, Moscow 117312, Russia}

\address[hse]{HSE University, Moscow}

\begin{abstract}
A new Highly-Granular time-of-flight Neutron Detector (HGND) is being developed and constructed to measure azimuthal neutron flow and neutron yields in nucleus-nucleus interactions in heavy-ion collisions with energies up to 4A GeV in the fixed target experiment BM@N at JINR.
Details of the detector design and results of performance studies for neutron identification and reconstruction are shown. Comparison of simulations for different options of the HGND layout at the BM@N is presented.
Several proposed methods of neutron reconstruction including machine learning and cluster methods are discussed.

\end{abstract}


\maketitle

\input{introduction}

\input{HGND_at_BMN}

\input{performance_studies}

\input{neutron_reconstruction}

\section{Summary}

The concept of new time-of-flight neutron detector with high granularity HGND has been developed for the BM@N experiment.
The new HGND detector allows to identify neutrons and to measure their energies from ion-ion collisions at BM@N energies.
The simulation results for the performance studies of the HGND detector at the BM@N for Bi+Bi @ 3.0A GeV are shown. 
The estimation of reconstructed neutron yieds during one month of BM@N operation will give about $1\times10^{9}$ neutrons.
Two methods of the neutron identification and measuring neutron energy are proposed and preliminary results of the efficiency and purity of these methods are shown.

\section*{Acknowledgments}
This work was carried out at the Institute for Nuclear Research, Russian Academy of Sciences, and supported by the Russian Scientific Foundation grant \textnumero 22-12-00132. 
The research leading to these results has received funding from the Basic Research Program at the
National Research University Higher School of Economics.
This research was supported in part through computational resources of HPC facilities at HSE University.

\bibliographystyle{elsarticle-num}
\bibliography{biblio}

\end{document}

%% file: introduction.tex
\section{Introduction}
\label{introduction}

The study of the equation of state (EoS) of dense nuclear matter is a key topic in relativistic nuclear physics.
The EoS that can be defined as a binding energy per nucleon relates the pressure, density, energy, temperature for symmetric matter and symmetry energy  components~\cite{Burgio:2020fom}.
The symmetric matter term of the EoS describes the matter in their ground state where protons and neutrons have equal contributions to the binding energy.
This term has been under extensive study and has the most stringent constraints from the vast set of experiments using observables such as meson yields, anisotropic flow measurements, etc.~\cite{E895:1999ldn, E895:2000maf, E895:2001axb, Sorensen:2023zkk}.
However the discrepancies in the existing results introduces the significant systematic uncertainty in the EoS constraints~\cite{Sorensen:2023zkk}.
New experimental data and state-of-the-art analysis techniques are required in order to address this issue.
The symmetry energy term of the EoS is responsible for the difference in isospin between neutron and proton parts of nuclear matter~\cite{Burgio:2020fom}. Study of this term is important for astrophysics because the mass and size of neutron stars and how they merge depend on the symmetry energy term in the EoS of high-density neutron matter. 
The only existing contraints on the symmetry energy comes from the ratio of neutron to proton anisotropic flow measured at FOPI/LAND experiment at 400 MeV per nucleon~\cite{FOPI:1993wdf, FOPI:1994xpn}.
One of the main problem of the symmetry energy study is the lack of experimental data in the high baryon density region (i.e. at beam energies higher than 1 GeV per nucleon).
Model study and comparison with the available experimental data suggests high sensitivity of the observables such as neutron to proton yields and ratios of neutron to proton anisotropic flow to the symmetry energy~\cite{Long:2024ggx}.

The BM@N (The Baryonic Matter at Nuclotron) provides excellent opportunities to study both terms of the EoS at high nuclear densities and obtain new experimental results~\cite{Senger:2022bzm}. First Xe+CsI data to study syimmetric term of EoS was already collected at 3.8A GeV in 2023.
A new highly granular time-of-flight neutron detector is being developed to measure neutron yields and flow at the BM@N for heavier colliding systems with ions up to 4A GeV.
The expected neutron flux consists of primary neutrons from the nucleus-nucleus interaction and background secondary neutrons.
The neutrons which are produced in inelastic nucleus-nucleus collisions have a continuous spectra of energy, rapidity and transverse momentum.
The estimated neutron flux at the BM@N is about 500 neutrons per second in the neutron detector acceptance (the details are discussed in Section~\ref{performance_studies}).
The neutron detector energy coverage ranges from 300 MeV to 4 GeV and the detector design requires a good detection efficiency in this range.
The time resolution of neutron detector cells is required to be at the level of 100-150 ps which is essential for robust neutron reconstruction.
In this article the simulations of Bi+Bi at 3A GeV reaction at the BM@N are used for the neutron detector performance studies.

%% file: HGND_at_BMN.tex
\section{The BM@N experiment at the NICA ion-accelerating complex}
\label{HGDN_at_BMN}

The BM@N (Baryonic Matter at Nuclotron) is the first experiment operational at
the Nuclotron/NICA ion-accelerating complex, dedicated to studying interactions of 
relativistic beams of heavy ions with fixed targets~\cite{BMN_EPJ} in the energy range that allows reaching high densities of baryonic matter~\cite{Cleymans}. The Nuclotron will provide the experiment with beams of a variety of particles, from protons to gold ions, with kinetic energy in the  range from 1 to $6\,{\rm AG}e{\rm V}$ for light ions with a $Z/A$ ratio of $\sim 0.5$ and up to $4.5\,{\rm AG}e{\rm V}$ for heavy ions with a $Z/A$ ratio of $\sim 0.4$.
At such energies, the density of nucleons in the fireball created by two colliding heavy nuclei is 3\,$-$\,4 times higher than the nuclear saturation density~\cite{Friman}. The primary goal of the experiment is to explore the QCD phase diagram in the region of high baryonic chemical potential, to search for the onset of critical phenomena, in particular, the conjectured critical end point, and to constrain the parameters of the equation of state (EoS) of high-density nuclear matter. In addition, the Nuclotron energies are high enough to study strange mesons and (multi)-strange hyperons produced in nucleus-nucleus collisions close to the kinematic threshold~\cite{NICAWhitePaper,BMN_CDR}.

\subsection{Neutron detector for the BM@N experiment}

\begin{figure}[!htp]
	\centering 
	\includegraphics[width=0.9\textwidth]{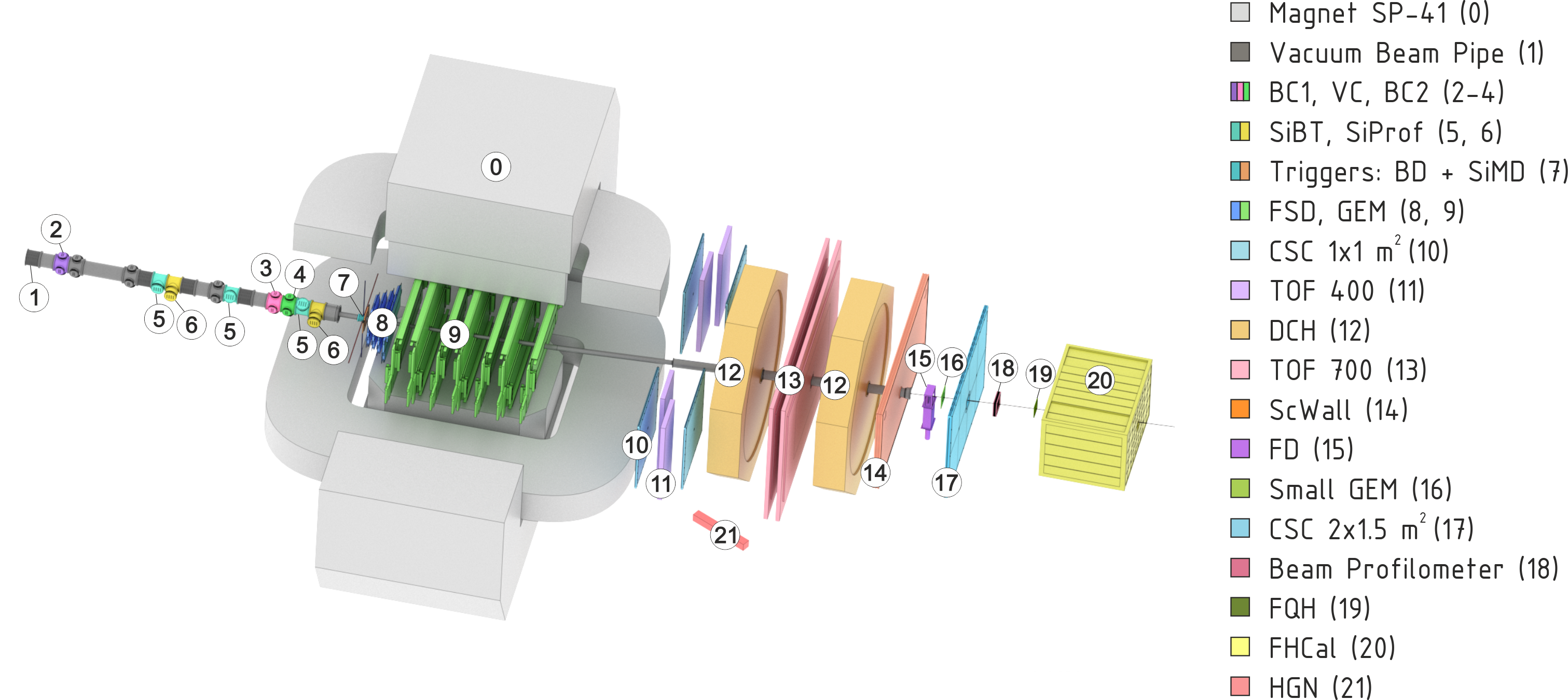}	
	\caption{Schematic view of the BM@N experiment. The legend has list of sub-detectors~\cite{afanasiev2024}.} 
	\label{fig:BMN}%
\end{figure}

The schematic view of the BM@N detector systems~\cite{afanasiev2024} is shown in Fig.~\ref{fig:BMN}. The setup comprises a dipole magnet along with several detector systems to monitor the beam, to identify produced charged particles, to measure their momentum, and to measure the collision centrality and reaction plane orientation of nucleus-nucleus collisions.
The beam detectors consist of three silicon beam trackers used to determine the trajectory of beam particles upstream the target. The Beam Counter (BC2) generates a starting signal. The tracking detectors consist of the central tracking system, based on seven planes of triple gas electron multipliers (GEM) measuring the momentum of charged particles in magnetic field. The system includes double-sided silicon micro-strip detectors (FSD) for high-precision vertex determination and track reconstruction. The outer tracking system containing six planes of Cathode Strip Chambers (CSC) enhances track parameter precision and efficiency. The particle identification detectors include Time-of-Flight (ToF) systems ToF 400 and ToF 700, based on multi-gap resistive plate chamber (mRPC) technology, enabling hadron (pi, K, p) and light nuclei separation with momentum up to a few GeV/c. The collision geometry detectors consist of the Forward Hadron Calorimeter (FHCal), the Forward Quartz Hodoscope (FQH) and the Scintillation Wall (ScWall) used to estimate centrality and reaction plane of the collision. The trigger system contains the barrel detector together with the group of beam detectors producing  trigger signal for the DAQ of the BM@N.

\begin{figure}[!htp]
	\centering 
	\includegraphics[width=1.0\textwidth]{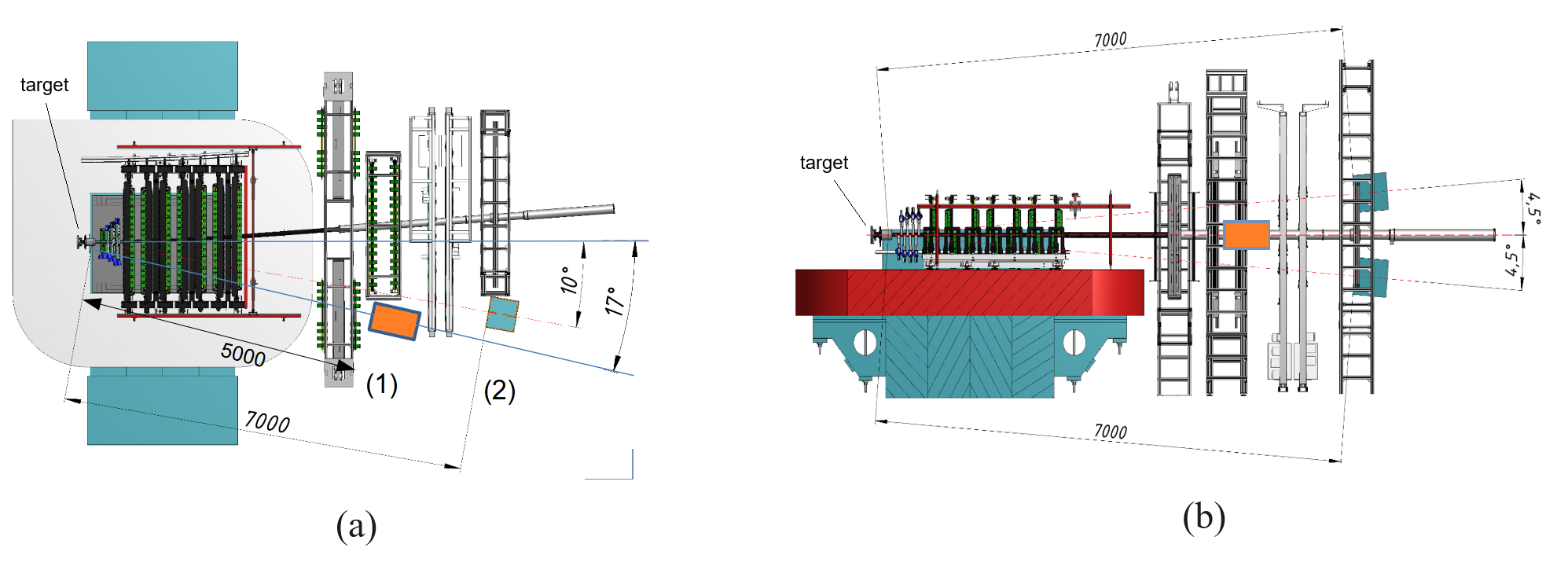}	
	\caption{Left (a): Top view of position options for the HGND at BM@N experimental area. Position 1 (red box) is for 16 layers single module option. Position 2 (blue boxes) is for "2-arms" option. Right (b): The side view of the single and "2-arms" options of the HGND at the BM@N.} 
	\label{fig:HGND_at_BMN_1_2}%
\end{figure}

Taking into account the available space to place the HGND at BM@N experimental area, two options are considered. The possible positions of the HGND are shown in  Fig.~\ref{fig:HGND_at_BMN_1_2}.  At position (1) the distance from the target is 5 m and at position (2) the distance is 7 m.
At these positions the angles of the HGND relative to the incident direction of the ion beam are 17 and 10 degrees, correspondingly.
In order to compensate the decreasing of detector acceptance at larger distance, the HGND positioned at 7 m is proposed to be divided into two parts ("arms") placed out of the horizontal plane. The side view of the HGND options at the BM@N is shown in Fig.~\ref{fig:HGND_at_BMN_1_2} (right).

\begin{figure}[!htp]
	\centering 
	\includegraphics[width=0.4\textwidth]{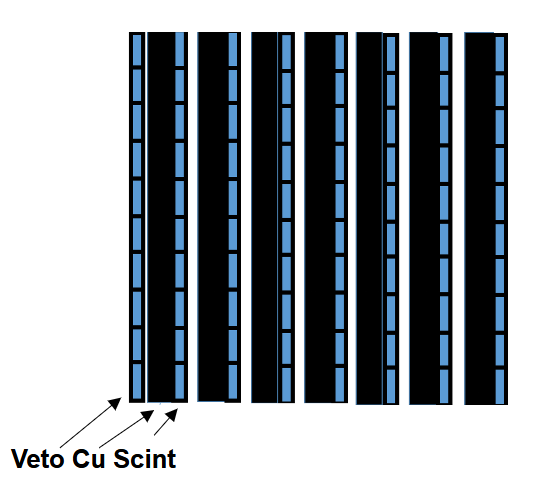}
        \includegraphics[width=0.3\textwidth]{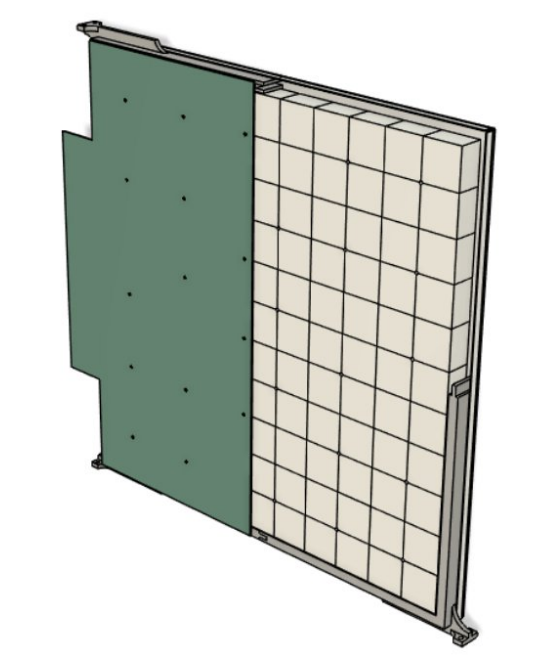}
	\caption{Left: schematic view of HGND structure. Right: The HGND active layer view.} 
	\label{fig:HGND_schematic_1}%
\end{figure}

\begin{figure}[!htp]
	\centering 
	\includegraphics[width=0.5\textwidth]{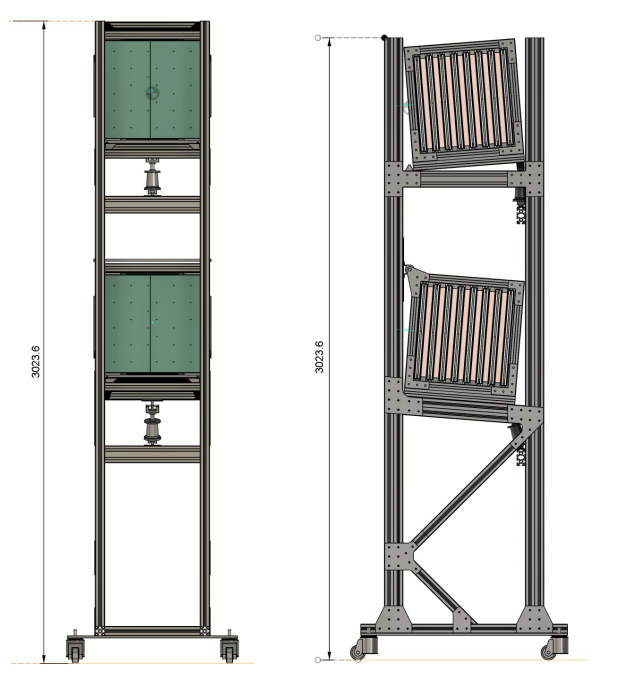}
        \includegraphics[width=0.38\textwidth]{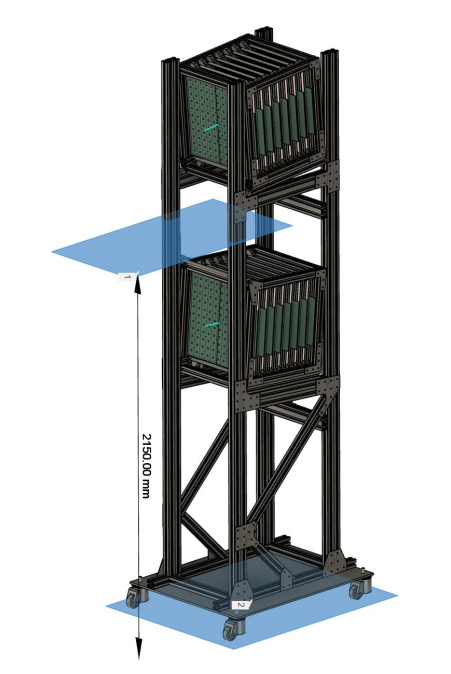}
	\caption{The schematic view of HGND support structure: the view from the beam point (left), the side view (center) and 3D view (right).} 
	\label{fig:HGND_schematic_support}%
\end{figure}

The HGND concept positioned at 5m from the target includes 16 alternating active layers with 121 scintillator cells grouped in an 11×11 array, with copper absorber plates of 3 cm thickness placed in between~\cite{guber2023}.
The first scintillation layer will be used as a VETO detector for charged particles.
As already mentioned, each module of the "2-arms" HGND positioned at 7m is shorter and consists of 8 scintillator layers with copper absorber in between.   The HGND module schematic view is shown in Fig.~\ref{fig:HGND_schematic_1} (left).
The array of scintillators is assembled in a common support frame (Fig.~\ref{fig:HGND_schematic_1}, right).
The cells are made up of $4\times4\times2.5~cm^{3}$ plastic scintillators based on polystyrene with additions of 1.5\% paraterphenyl
and 0.01\% POPOP. This plastic scintillator, with a light decay time of 3.9$\pm$ 0.7 ns, is produced at JINR.
Each active layer of the HGND is assemled in a 3D-printed PETG light-tight frame with 121 scintillators arranged in $11\times11$ matix.
The frame is capped with two 1.5 mm aluminum sheets.
Two PCBs with 55 and 66 SiPMs, respectively, are mounted on the downstream side of the scintillators, providing light readout and hosting preamplifiers and LVDS comparators, which generate ToT signals for the readout schematics.
On the upstream side of the scintillators, a PCB with 121 LEDs is mounted, providing full-circuit calibration capabilities.
Both HGND modules are placed on the common support structure as shown in Fig.~\ref{fig:HGND_schematic_support}.

\subsection{The read-out electronics of the HGND}

The HGND read-out is based on detector-mounted electronics concept without signal cables.
The readout boards with FPGAs are mounted on both lateral sides of the HGND connected to the PCB via PCIe.
The readout board with a 100 ps FPGA-based TDC (Time to Digital Converter) is currently under development~\cite{Finogeev_2024}. The HGND will consist of eight such readout boards, each comprising three Kintex 7 FPGAs for reading out 252 channels. The TDC operates on the standard LVDS 4x asynchronous oversampling and is synchronized with the experiment timestamp using the White Rabbit link.
The measurements show channel precision on the level of 40 ps.

%% file: performance_studies.tex
\section{Performance studies for the HGND detector}
\label{performance_studies}

The HGND performance was studied for two setup positions in the BM@N experimental area.
These simulation studies was performed with Geant4~\cite{Geant4} based package bmnroot~\cite{bmnroot}. The DCM-QGSM-SMM~\cite{Baznat_2020_DCM_SMM} generator was used for nucleus-nucleus collision simulation.

\begin{figure}[!htp]
	\centering 
	\includegraphics[width=1.0\textwidth]{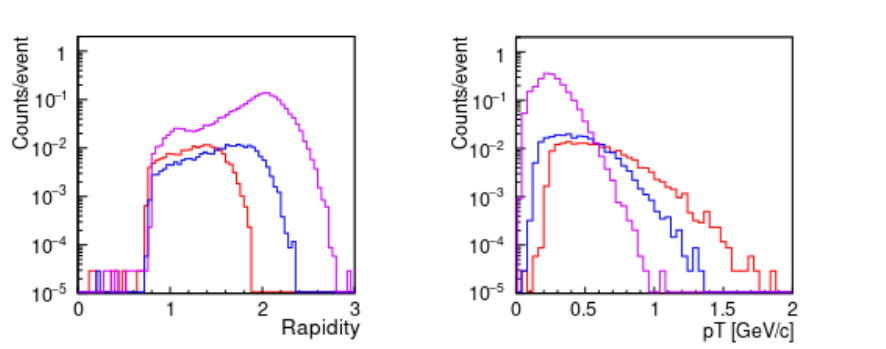}	
	\caption{Left: neutron rapidity spectra for different positions of the HGND. Right: neutron transverse momentum spectra. Different colors represent different positions and angles of the HGND: red - 17 deg at 5 m, blue - 10 deg at 7 m and magenta - 4.7 deg at 7 m. } 
	\label{fig:performance_1}%
\end{figure}

The neutron rapidity spectra for different HGND setups are shown in Fig.~\ref{fig:performance_1} (left) and the transverse momentum spectra are presented in Fig.~\ref{fig:performance_1} (right).

The rapidity spectra for the HGND cover region of mid-rapidity ($\sim$1.06) and extend up to 2.4 on the left tail. The angle is chosen to keep the rapidity spectrum wide as well as avoid the spectator neutrons from the collisions at angles below 4.7 degrees.

\subsection{Primary and background neutrons on the HGND}

\begin{figure}[!htp]
	\centering 
	\includegraphics[width=1.0\textwidth]{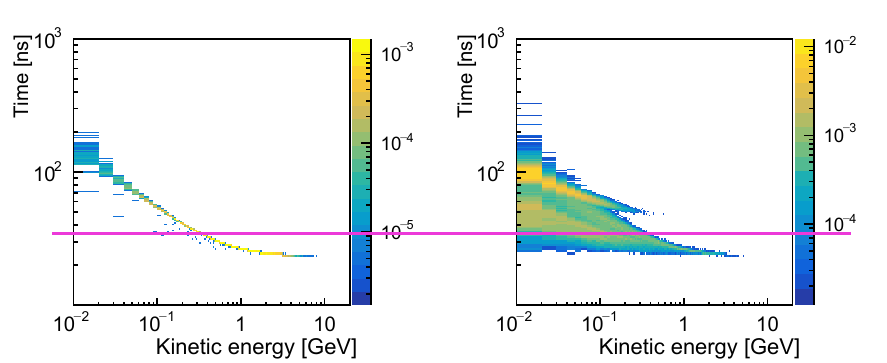}
	\caption{Left: primary neutrons at the HGND surface. Right: background neutron on all surfaces of the HGND. The line (magenta) represents a 35ns time cut.} 
	\label{fig:particles_at_HGND_1}
\end{figure}

\begin{figure}[!htp]
	\centering 
	\includegraphics[width=0.6\textwidth]{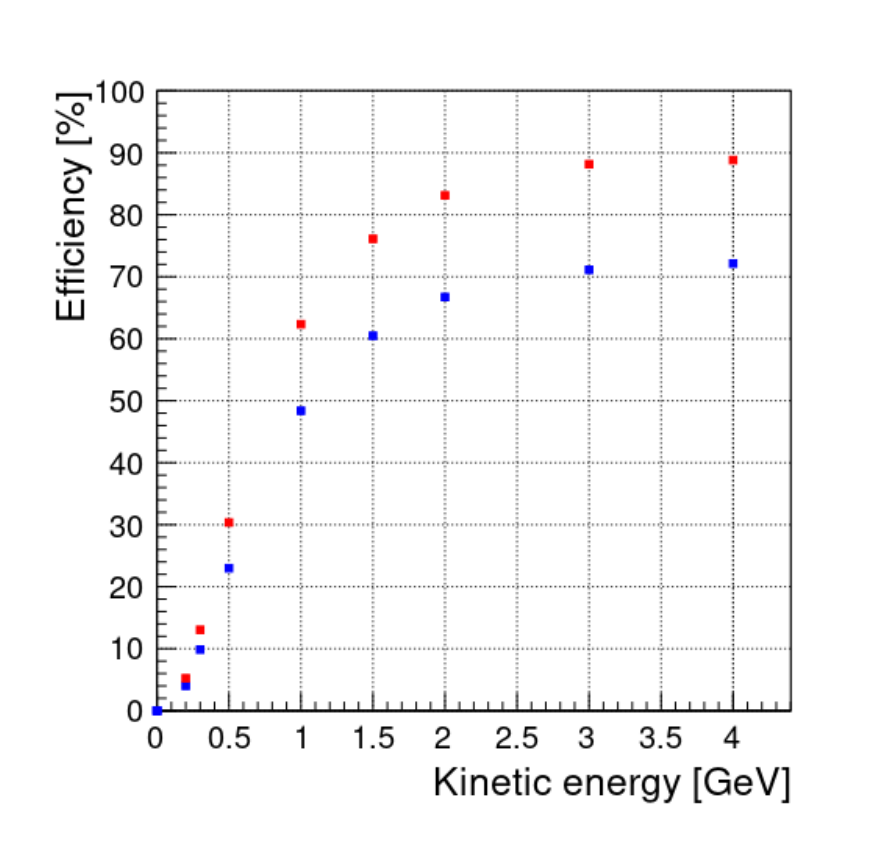}
	\caption{The result of simulation for neutron detection efficiency study. Red dots: single HGND option, blue dots: "2-arms" HGND option.} 
	\label{fig:neutrons_eff}
\end{figure}

\begin{figure}[!htp]
	\centering 
\includegraphics[width=1.0\textwidth]{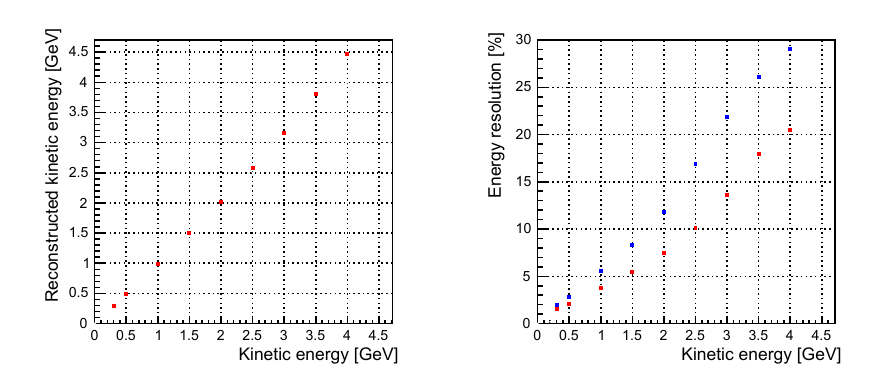}
	\caption{Left: Mean value of energy reconstructed for different energy of incident neutrons for 7m HGND position from the target. Right: energy resolution of neutron reconstruction. Red dots are for 7m distance and blue dots are for 5m distance from the target to the HGND.} 
	\label{fig:neutrons_lin_res}
\end{figure}

\begin{figure}[!htp]
	\centering 
	\includegraphics[width=0.4\textwidth]{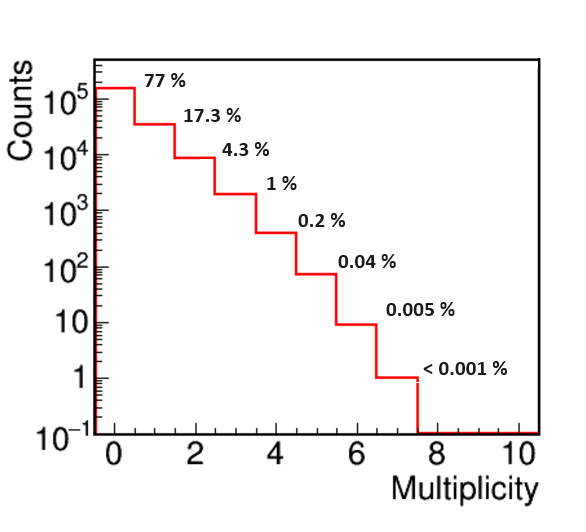}
        \includegraphics[width=0.48\textwidth]{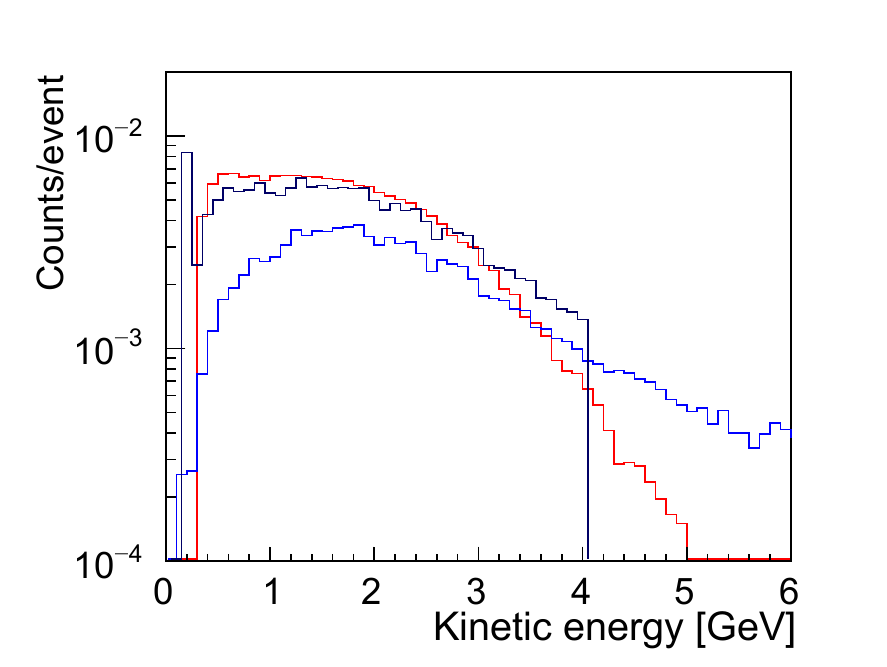}
	\caption{Left: Neutron multiplicity for Bi+Bi @ 3.8A GeV for 2-arms HGND option. Right: single neutron event energy reconstruction: red line is the spectrum of incoming neutrons, blue line is the reconstructed spectrum and black line is the reconstructed energy corrected with neutron detection efficiency.} 
	\label{fig:neutrons_mul_spectrum}
\end{figure}

The performance studies for the HGND optimization has been done for Bi+Bi reaction at 3.0A GeV. The primary and background neutrons time-of-flight spectra for the 10 deg position are shown in Fig.~\ref{fig:particles_at_HGND_1}. 
The time cut of 35 ns shown in the pictures significantly (about 6 times) reduces the background neutrons yield. This cut reduces total statistics of primary neutrons by 8\% only and allows to register primary neutrons from about 300 MeV.
In order to study the neutron detection efficiency of the HGND the simulation of response of the detector to single energy neutrons has been done.
The probability of the primary neutrons entering the HGND acceptance to be registered is shown in Fig.~\ref{fig:neutrons_eff} as a function of its kinetic energy.
The neutron detection efficiency is calculated for different energies of neutrons and for two options of the HGND construction. The difference for 1 GeV neurons for single HGND and "2-arms" HGND is about 15\% only.
The time resolution for the HGND scintillation cell was measured on the electron beam at "Pakhra" synchrotron (FIAN, Troitsk) and has a value of about 130 ps with the beam centered on the cell and about 150 ps in average for full cell surface~\cite{guber2023TimeResCells}.
The energy of neutron is calculated using the time-of-flight method based on the hit in the HGND cell with minimum time. 
The linearity of reconstructed energy with the HGND for different energies of neutrons as well as energy measurement resolution is shown in Fig.~\ref{fig:neutrons_lin_res} for 150 
ps intrinsic energy resolution.
The multiplicity of primary neutrons at the HGND surface has been studied based on simulations. The neutron multiplicity is shown in Fig.~\ref{fig:neutrons_mul_spectrum} (left).
The neutron yield estimation is based on the results of multiplicity simulation. The single neutron events are about 17.3\% of total events on the HGND surface and about 76\% of total neutron events.
Taking into account the BM@N operation efficiency ($1\times10^{6}$ ions per spill, 50\% duty factor and 70\% efficiency of Nuclotron), 2\% interaction length of target and mean efficiency of the HGND of 50\% one gets about $1.2\times10^{9}$ single neutron events reconstructed during one month of BM@N operation.
If the multiple neutron events will be recognized as well the statistics of neutron detection will rise up to $1.5\times10^{9}$ for one month of BM@N data taking.
The reconstructed spectrum of single neutrons is shown in Fig.~\ref{fig:neutrons_mul_spectrum} (right).
The red line in this picture is a spectrum of the incoming neutrons at the HGND surface.
The blue line is the reconstructed energy of neutrons with time-of-flight method.
The black line represents reconstructed neutron spectrum corrected with the efficiency (Fig. ~\ref{fig:neutrons_eff}).

\begin{figure}[!htp]
	\centering 
	\includegraphics[width=0.6\textwidth]{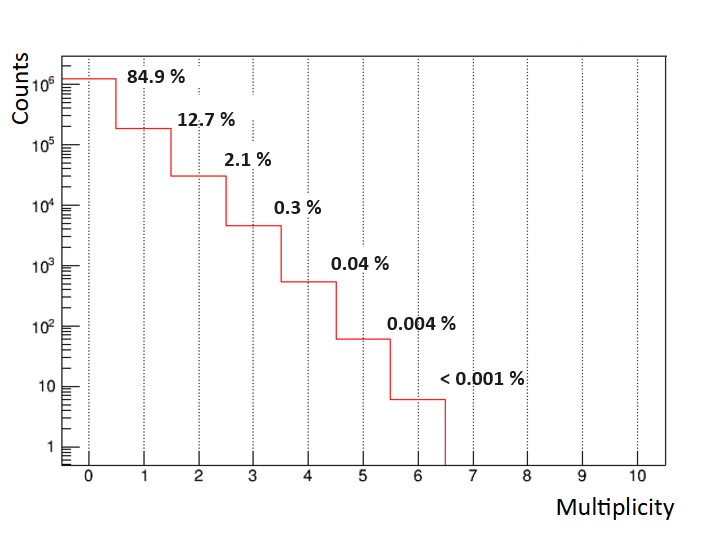}
	\caption{Neutron multiplicity for Xe+CsI @ 3A GeV for "2-arms" HGND option.} 
	\label{fig:neutrons_mul_Xe}
\end{figure}

The first data with the HGND in the BM@N experiment will be taken with Xe ion beam. In order to estimate the neutron statistics for the first period of BM@N operations with Xe beam the simulation for Xe+CsI @ 3.8A GeV with DCM-QGSM-SMM generator is done. The result of neutron multiplicity calculation is presented in Fig.~\ref{fig:neutrons_mul_Xe}.
The number of single neutron events for Xe+CsI @ 3.8A GeV is about 12.7\% of total events at the HGND surface. Taking into account the BM@N operation efficiency (as above) the statistics of neutron detection is about $0.9\times10^{9}$ single neutron events per month.

%% file: neutron_reconstruction.tex
\section{Concepts of neutron identification and neutron energy reconstruction}
The measurement of neutron yields and flow  requires the reconstruction of neutrons. This task involves the identification of neutrons produced in the reaction in the presence of background and the reconstruction of neutron kinematics.
Two different methodological approaches are being developed.
The first one is a cluster method and the second one is a machine learning based approach.
The use of two different techniques allows cross-checking of the results.

\begin{figure}[h!]
\begin{center}
    \includegraphics[width=0.7\linewidth]{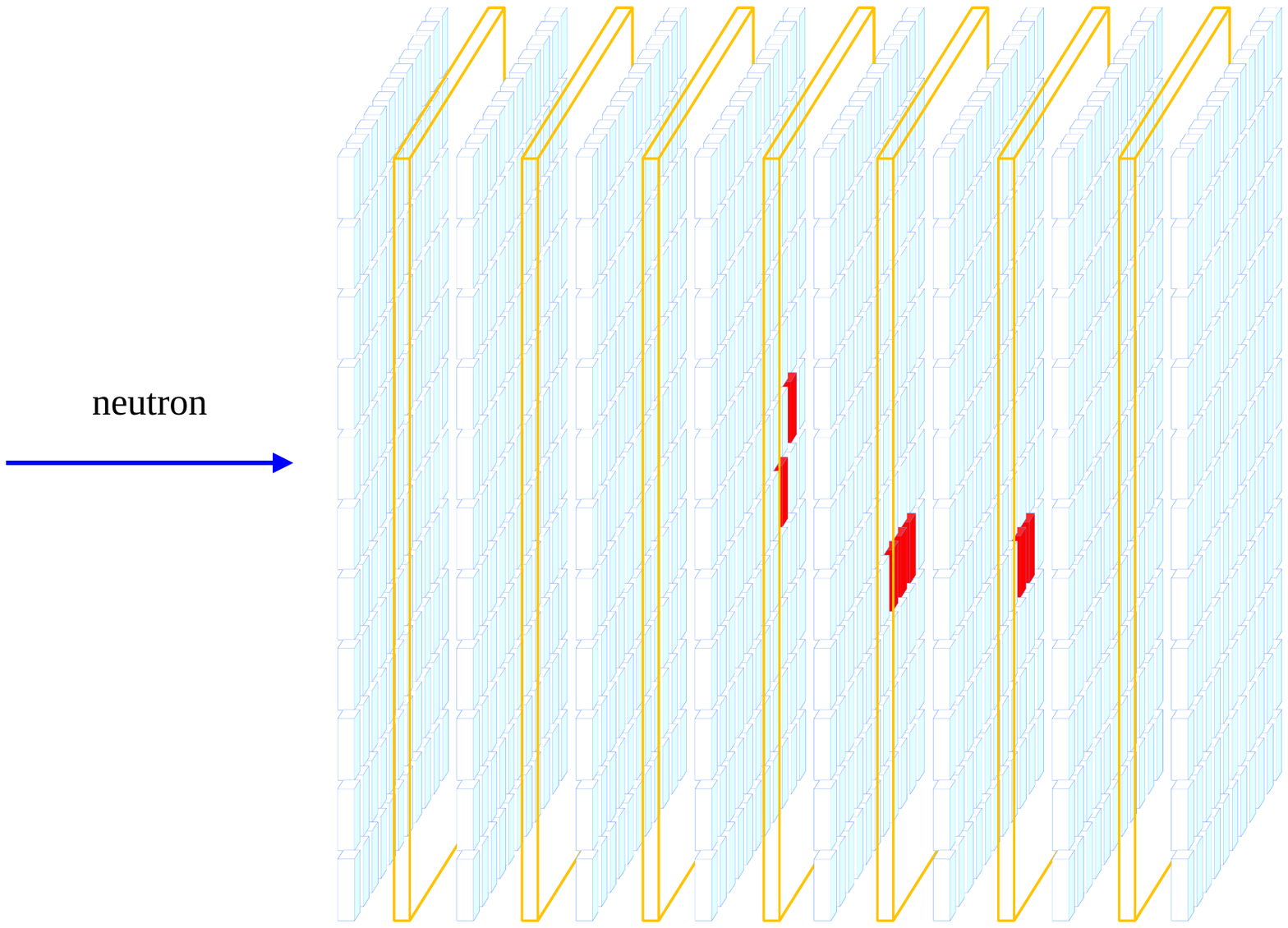}
    \caption{Example of an event with a neutron (kinetic energy of 1.58 GeV) hitting the HGND. Red boxes represent the fired cells.}
    \label{fig:example}
\end{center}
  \end{figure}

Both methods were tested with  simulated Bi+Bi reaction at 3.0A GeV. One can see an example of such an event on Figure \ref{fig:example}.
Only cells with an energy deposition above the threshold ($E_{dep}>3MeV$), which corresponds approximately to half of the energy loss of the minimum ionising particle (MIP), and a hit time \(t_{hit}<35ns\), corresponding to the time of flight of neutrons with an energy $E_{kin}>300 MeV$,  are used for analysis. Additionally, Gaussian time smearing with standard deviation of $150~ps$ is applied to each hit.

\subsection{Cluster method}
{
The cluster method is the traditional approach to analysing data from highly granular detectors. It is a three-step process: first, merging multiple cells into clusters; second, determining the type of particle that produced the cluster; and third, determining the kinematic parameters of the particle.

The clustering of the fired cells is based on their position and time of flight. Neighbouring cells with close time stamps are grouped into clusters. Each cluster contains the cell with the highest velocity, defined as $v=d/t$, where $d$ is the distance from the cell to the target and $t$ is the time of flight. The cell with the highest $v$ is designated as the "head of the cluster". Sub-clusters with close speed are combined into a larger cluster.

In order to select the clusters produced by neutrons, the following criteria are applied:
\begin{itemize}
\item The clusters containing cells of the veto layer are rejected. This criterion results in the removal of the charged particles.
\item Clusters containing 1st layer cells are rejected. This cut suppresses $\gamma$. 
\item Additional suppression of $\gamma$ and light charged particles is achieved by a velocity cut: $v<c$
\end{itemize}

The kinetic energy of the neutron is calculated from the time of flight as follows: 
\begin{center}
$E_{tof} = m_{n} \cdot (\sqrt{\frac{1}{1-(\frac{v}{c})^2}}-1)$,\\
\end{center}
where $m$ is the mass of neutron and $v$ is the velocity of the cluster. The expected number of fired cells, sub-clusters and total deposited energy depend on the kinetic energy of a neutron. This allows for the verification of whether the cluster could have been produced by a neutron with the reconstructed energy $E_{tof}$. Furthermore, these parameters can also be used to distinguish between single and multiple neutron events.
}

\subsection{Machine Learning method}
{
    The machine learning approach employs Graph Neural Networks (GNNs)~\cite{GNN} to improve neutron reconstruction performance by examining the geometric structures of events. 

    Each node represents a hit, characterized by key observables, including hit coordinates and energy deposition. Additionally, a global event node is introduced to encapsulate four parameters related to the total energy and timing characteristics of the event. The constructed event graphs are then divided into training and testing datasets using a 50/50 split to ensure robust model evaluation.

    The reconstruction procedure is divided into two primary components: the identification of neutron events and the energy regression of the fastest neutron within those events. Two distinct GNN models were developed and trained independently: the Classification GNN, which classifies events based on neutron contributions, and the Energy Regression GNN, which predicts the energy of the fastest neutron associated with each event. Notably, the Energy Regression GNN is trained exclusively on events identified as having neutron contributions.

}

\subsection{Performance of neutron reconstruction}
{
    The preliminary performance of both proposed methods was evaluated through a simplified reconstruction scenario focusing on the reconstruction of a single neutron. 
    For neutron multiplicities greater than one, only the fastest neutron was considered. 
    Two key performance metrics for neutron reconstruction were evaluated: reconstruction efficiency and the purity of the reconstructed signal. Neutron reconstruction efficiency is defined as the ratio of the number of successfully reconstructed neutron events that contain at least one true neutron to the total number of events that include at least one neutron. Signal purity is defined as the ratio of the number of reconstructed neutron events containing at least one true neutron to the total number of reconstructed neutron events.
    Figure \ref{fig:eff_reco} illustrates the dependence of neutron reconstruction efficiency on neutron kinetic energy, maintaining a fixed signal purity of 0.7.

  

  \begin{figure}[H]
    \centering
    \includegraphics[width=0.9\linewidth]{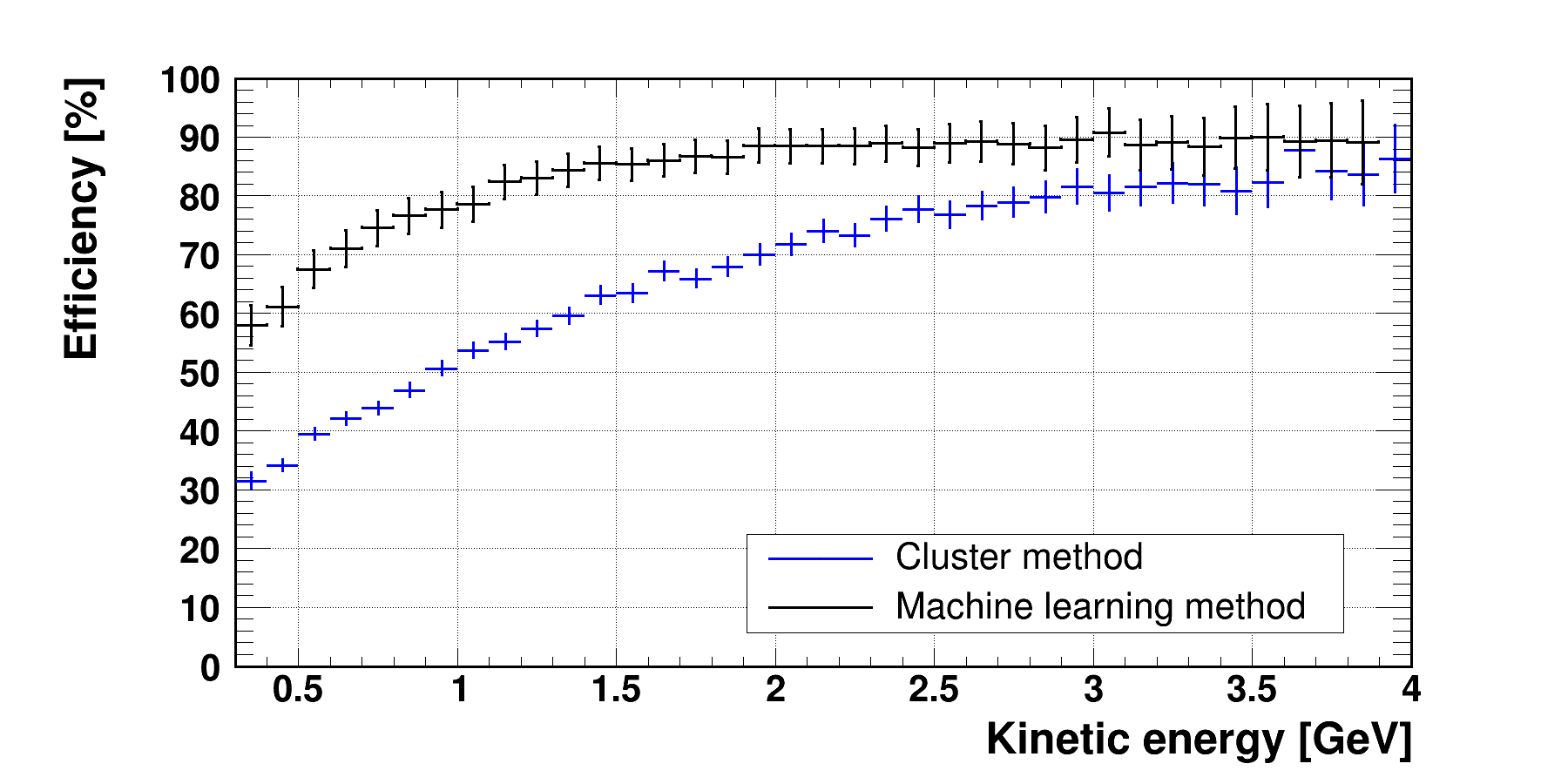}
    \caption{Efficiency of neutron reconstruction at purity=0.7}
    \label{fig:eff_reco}
  \end{figure}

    Figure \ref{fig:esim_erec} illustrates the energy reconstruction performance for MC-truth neutron events. 
    The results of the comparison between the two methods are presented for $7\cdot10^4$ events. 
    It can be seen that the results are satisfactory for both cases.
    
    \begin{figure}[H]
      \centering
      \includegraphics[width=0.9\linewidth]{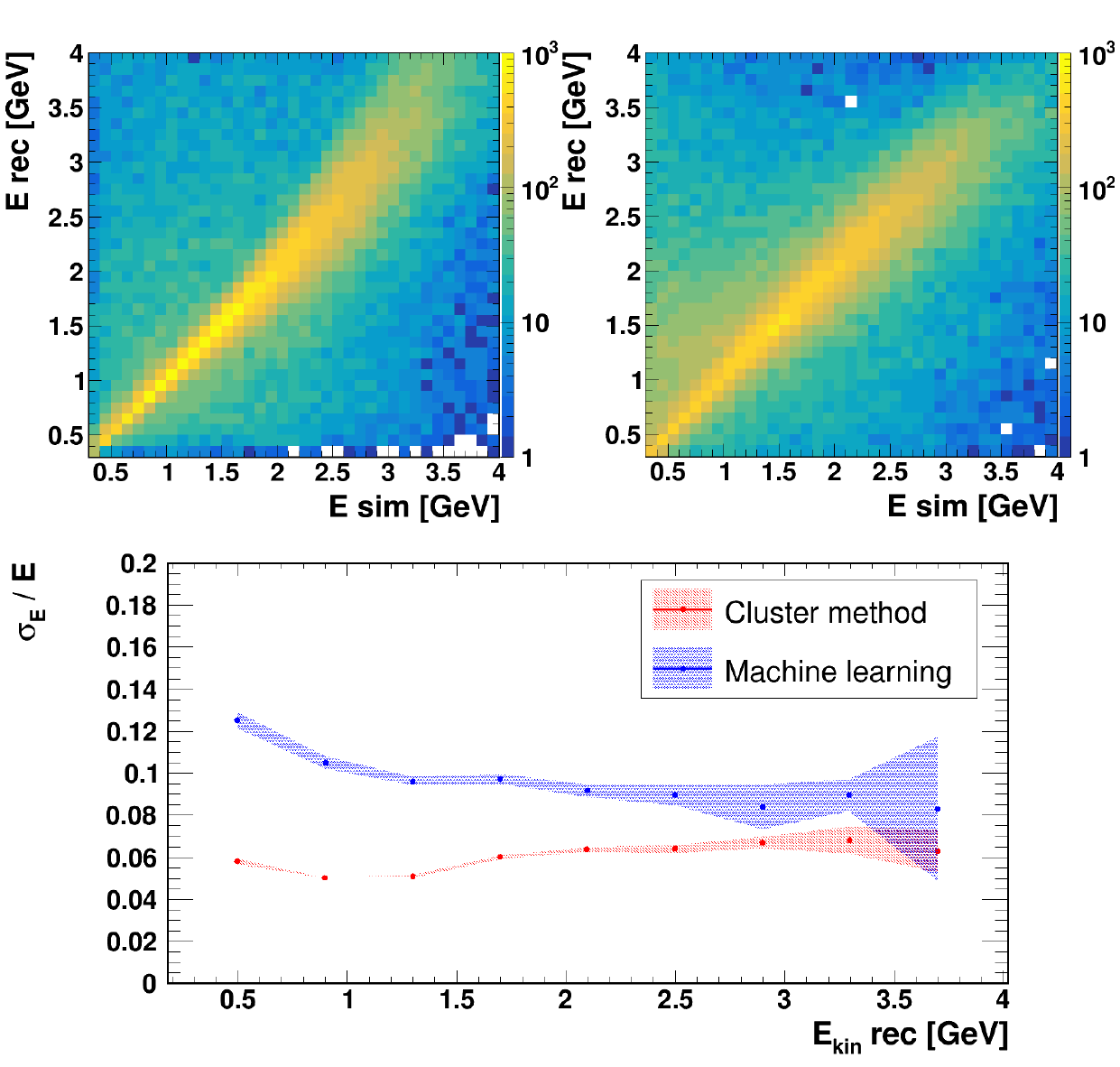}
      \caption{Dependence of reconstructed kinetic energy of the neutron on its simulated energy using the cluster method (top left) and the machine learning technique (top right). The energy resolution of both methods is shown on the bottom plot.}
      \label{fig:esim_erec}
    \end{figure}

    The clustering method yields results that are well understood and predictable.
    At low energies it provides better resolution than the ML approach.
    However, at higher energies, it gives a systematic shift in the energy determination.
    This effect depends on the time resolution and can be negotiated via non-linear error propagation in the time-of-flight technique.
    
    The ML-based neutron reconstruction approach yielded promising results across various performance metrics. 
    High signal efficiency was achieved for energies around 1 GeV and above, with a well-defined linear correlation observed up to energy levels of 3-4 GeV. 
    Additionally, the model effectively compensates for time-of-flight overestimations within the energy range of 2-4 GeV.


}